\documentclass[submitting]{nst}

\usepackage{subfigure,dcolumn}
\usepackage{epstopdf}
\usepackage[version=4]{mhchem}
\usepackage{upgreek}

\begin{document}

\title{Performance study of the Highly Granular Neutron Detector prototype in the BM@N experiment}

\author{A.\,Zubankov}
\email[Corresponding author, ]{zubankov@inr.ru}
\affiliation{Institute for Nuclear Research of the Russian Academy of Sciences, 117312, Prospekt 60-letiya Oktyabrya 7a, Moscow, Russia}
\affiliation{National Research Nuclear University MEPhI, 115409, Kashirskoe sh. 31, Moscow, Russia}

\author{S.\,Afanasiev}
\affiliation{Joint Institute for Nuclear Research, 141980, Joliot-Curie St. 6, Dubna, Russia}

\author{M.\,Golubeva}
\affiliation{Institute for Nuclear Research of the Russian Academy of Sciences, 117312, Prospekt 60-letiya Oktyabrya 7a, Moscow, Russia}

\author{F.\,Guber}
\affiliation{Institute for Nuclear Research of the Russian Academy of Sciences, 117312, Prospekt 60-letiya Oktyabrya 7a, Moscow, Russia}

\author{A.\,Ivashkin}
\affiliation{Institute for Nuclear Research of the Russian Academy of Sciences, 117312, Prospekt 60-letiya Oktyabrya 7a, Moscow, Russia}

\author{N.\,Karpushkin}
\affiliation{Institute for Nuclear Research of the Russian Academy of Sciences, 117312, Prospekt 60-letiya Oktyabrya 7a, Moscow, Russia}
\affiliation{Joint Institute for Nuclear Research, 141980, Joliot-Curie St. 6, Dubna, Russia}

\author{O.\,Kutinova}
\affiliation{Joint Institute for Nuclear Research, 141980, Joliot-Curie St. 6, Dubna, Russia}

\author{D.\,Lyapin}
\affiliation{Institute for Nuclear Research of the Russian Academy of Sciences, 117312, Prospekt 60-letiya Oktyabrya 7a, Moscow, Russia}
\affiliation{National Research Nuclear University MEPhI, 115409, Kashirskoe sh. 31, Moscow, Russia}

\author{A.\,Makhnev}
\affiliation{Institute for Nuclear Research of the Russian Academy of Sciences, 117312, Prospekt 60-letiya Oktyabrya 7a, Moscow, Russia}

\author{S.\,Morozov}
\affiliation{Institute for Nuclear Research of the Russian Academy of Sciences, 117312, Prospekt 60-letiya Oktyabrya 7a, Moscow, Russia}
\affiliation{Joint Institute for Nuclear Research, 141980, Joliot-Curie St. 6, Dubna, Russia}

\author{P.\,Parfenov}
\affiliation{Institute for Nuclear Research of the Russian Academy of Sciences, 117312, Prospekt 60-letiya Oktyabrya 7a, Moscow, Russia}
\affiliation{National Research Nuclear University MEPhI, 115409, Kashirskoe sh. 31, Moscow, Russia}
\affiliation{Joint Institute for Nuclear Research, 141980, Joliot-Curie St. 6, Dubna, Russia}

\author{I.\,Pshenichnov}
\affiliation{Institute for Nuclear Research of the Russian Academy of Sciences, 117312, Prospekt 60-letiya Oktyabrya 7a, Moscow, Russia}
\affiliation{Moscow Institute of Physics and Technology, 141701, Institutskiy per. 9, Dolgoprudny, Russia}

\author{D.\,Sakulin}
\affiliation{Joint Institute for Nuclear Research, 141980, Joliot-Curie St. 6, Dubna, Russia}

\author{S.\,Savenkov}
\affiliation{Institute for Nuclear Research of the Russian Academy of Sciences, 117312, Prospekt 60-letiya Oktyabrya 7a, Moscow, Russia}
\affiliation{Moscow Institute of Physics and Technology, 141701, Institutskiy per. 9, Dolgoprudny, Russia}

\author{A.\,Shabanov}
\affiliation{Institute for Nuclear Research of the Russian Academy of Sciences, 117312, Prospekt 60-letiya Oktyabrya 7a, Moscow, Russia}
\affiliation{Moscow Institute of Physics and Technology, 141701, Institutskiy per. 9, Dolgoprudny, Russia}

\author{E.\,Sukhov}
\affiliation{Joint Institute for Nuclear Research, 141980, Joliot-Curie St. 6, Dubna, Russia}

\author{A.\,Svetlichnyi}
\affiliation{Institute for Nuclear Research of the Russian Academy of Sciences, 117312, Prospekt 60-letiya Oktyabrya 7a, Moscow, Russia}
\affiliation{Moscow Institute of Physics and Technology, 141701, Institutskiy per. 9, Dolgoprudny, Russia}

\author{G.\,Taer}
\affiliation{Kurchatov Institute, NRC, 123182, Akademika Kurchatova pl. 1, Moscow, Russia}

\author{V.\,Ustinov}
\affiliation{Joint Institute for Nuclear Research, 141980, Joliot-Curie St. 6, Dubna, Russia}

\begin{abstract}
The time-of-flight Highly Granular Neutron Detector (HGND) with a multilayer longitudinal structure of interleaved absorber and scintillator plates, high transverse granularity and a time resolution of about 150~ps is currently under development.
The detector is designed to identify neutrons produced in nucleus-nucleus collisions and measure neutron kinetic energies of 0.3--4~GeV by the time-of-flight method in the BM@N experiment at the NICA accelerator complex at JINR.
In order to validate the concept of the full-scale HGND, a compact HGND prototype was first designed and built, and its performance was studied in the BM@N experiment. The acceptance of the HGND prototype and the detection efficiency of forward neutrons emitted in hadronic fragmentation and electromagnetic dissociation (EMD) of 3.8A~GeV $^{124}$Xe projectiles interacting with a CsI target were calculated by means of the DCM-QGSM-SMM and RELDIS models, respectively.
The energy distributions of forward spectator neutrons and neutrons from the EMD were measured and compared with the simulations.
The developed methods will be used to calibrate the full-scale HGND and to study its efficiency.
\end{abstract}


\maketitle

\nolinenumbers 

\section{Introduction}
\label{sec:intro}

The Baryonic Matter at Nuclotron (BM@N) is the first fixed target experiment at the NICA accelerator complex~\cite{bib:Afanasiev2024}. 
The research program of the BM@N is aimed at studying the equation of state (EoS) of nuclear matter at high baryon densities achieved in nucleus-nucleus collisions with kinetic energies of projectile nuclei up to 4.65$A$~GeV~\cite{bib:Mamaev2023}. 
The EoS relates the pressure, density and temperature achieved in a nucleus-nucleus collision event. 
The EoS includes the symmetry energy term to characterize the isospin asymmetry of nuclear matter, which is crucial for understanding the properties of astrophysical objects such as neutron stars~\cite{bib:Steiner2005}. 
It was shown~\cite{bib:Long2024ggx}, that the neutron to proton ratio in terms of yields and directed flow is sensitive to the symmetry energy contribution to the EoS of high-density nuclear matter.
The BM@N detector system has the capability to measure the anisotropic flow of protons using a hybrid tracking system together with TOF and FHCal detectors~\cite{bib:Mamaev2024bhl}. 
However, to measure the yields and directed flow of neutrons, additional detectors and identification methods are necessary.

Up to now, only two neutron detectors have been developed: LAND~\cite{bib:Pawlowski2023} and NeuLAND~\cite{bib:Boretzky2021}. Both were developed and constructed at GSI in Darmstadt, Germany to measure neutrons from heavy-ion collisions at beam energies below 1A~GeV.
The LAND consists of 10 mutually perpendicular layers with 20 individual long scintillation detectors in each layer.
Each individual detector consists of 10 alternating layers of plastic scintillation and iron plates with a thickness of 0.5 cm and has a size of $10\times10\times200$~cm$^3$, the light from two ends of which is read by two photomultiplier tubes.
The full geometric size of the LAND detector is $200\times200\times100$~cm$^3$.
A typical resolution of time measurements with the LAND is reported to be 250~ps.

A new large area neutron detector NeuLAND~\cite{bib:Boretzky2021} has been designed and partially constructed to study the properties of nuclear matter produced in collisions of radioactive nuclei, within the R3B project at the FAIR accelerator facility at GSI.
In contrast to LAND, NeuLAND consists only of active layers of plastic scintillation detectors with a size of $250\times5\times5$~cm$^3$ each.
The light from the opposite ends of the detectors is read by two photomultiplier tubes.
NeuLAND is composed of 60 active layers with 50 scintillation detectors in each layer and has a total size of $250\times250\times300$~cm$^3$.
The time resolution of NeuLAND is estimated to be 150~ps, the spatial resolution of the first interaction point of the neutron is about 1.5 cm, and the efficiency of detecting each neutron in an event is better than 95\% for kinetic energies of 400 -- 1000~MeV.

Taking into account the significantly higher energy (up to 4~GeV) of the neutrons produced in nucleus-nucleus collisions in the BM@N  experiment and the intense development of hadron showers in the detector volume at these energies, it is impossible to use large-size neutron detectors like LAND and NeuLAND.
Therefore, a new concept of Highly Granular Neutron Detector (HGND) has been proposed. 
Instead of the long scintillator plates used in the LAND and NeuLAND, it is proposed to use the highly granular transverse structure of the active elements of the detector to identify neutrons and measure their energy with the time-of-flight method with good resolution.
At the same time, the multilayer longitudinal sampling makes it possible to achieve a high efficiency of neutron detection~\cite{bib:Guber2024}.
It was proposed to build the active layers of the HGND as ($11\times11$) arrays of small scintillator cells with individual light readout by silicon photomultipliers (SiPM).
Each cell is expected to provide a time resolution of $\sim$150~ps. Copper absorbers placed between the active layers will provide a rather high probability of neutron interaction with the HGND.

Besides the studies of neutron yields and flow in hadronic interactions of colliding nuclei, it is also interesting to study the electromagnetic dissociation (EMD) of beam nuclei in ultraperipheral collisions (UPC) resulting in the emission of forward neutrons. 
The EMD is usually represented by the emission of a single or very few neutrons, producing a single residual nucleus~\cite{bib:Pshenichnov2024}.

In the present work, the design and performance of a compact prototype of the HGND are presented. 
The capability of the HGND prototype to identify forward neutrons produced in nuclear fragmentation and electromagnetic dissociation of 3.8A~GeV $^{124}$Xe projectiles on a CsI target and reconstruct the neutron energies was studied in the BM@N experiment.

The paper is organized as follows. 
In Sec.~\ref{sec:hgnd}, the design of the HGND prototype is presented.
In Sec.~\ref{sec:analysis} the event selection criteria and procedures to reconstruct the kinetic energy of neutrons are described. 
After that, in Sec.~\ref{sec:sim}, the energy spectra of neutrons are compared with Monte Carlo modeling, and the acceptance and efficiency of neutron detection by the HGND prototype are estimated. 
The results are summarized in conclusion, Sec.~\ref{sec:concl}.

\section{Highly Granular Neutron Detector (HGND) and its prototype}
\label{sec:hgnd}

\subsection{Concept of the HGND}
The original concept of the full-scale HGND consisting of 16 alternating active layers of 121 plastic scintillation detectors grouped in the $11 \times 11$ array, with copper absorber plates in between, has been presented in detail in~\cite{bib:Guber2024}.
The absorber plates (44 $\times$ 44 $\times$ 3 cm$^{3}$) and the scintillator layers (44 $\times$ 44 $\times$ 2.5 cm$^{3}$) will be mounted on a common support frame.
The total length of the HGND is about 1~m and corresponds to $\sim$3  nuclear interaction lengths $\lambda_{\rm int}$.
The first scintillation layer of the HGND will be used as a VETO detector for charged particles. 
The scintillator layers will be assembled from 4 $\times$ 4 $\times$ 2.5 cm$^{3}$ plastic scintillators (cells) based on polystyrene with additions of 1.5\% paraterphenyl and 0.01\% POPOP produced at JINR.  
The light read-out from the center of entrance surface of each cell is conducted with EQR15 11-6060D-S silicon photomultipliers~\cite{bib:Eqr15_2025, bib:Guber2024b}  placed on the printed circuit board (PCB) that hosts preamplifier and Low-Voltage Differential Signal (LVDS) comparators, generating Time over Threshold (ToT)~\cite{bib:KARPUSHKIN2024169739} signals for the read-out schematics.
The methods to identify neutrons in high multiplicity neutron events in the presence of the background from charged particles and photons are currently under development.
More technical details of the HGND design as well as results of the simulations can be found in~\cite{bib:Guber2024}.

\subsection{Design of the HGND prototype}

\begin{figure*}[htb!]
\centering
\includegraphics[width=.9\textwidth]{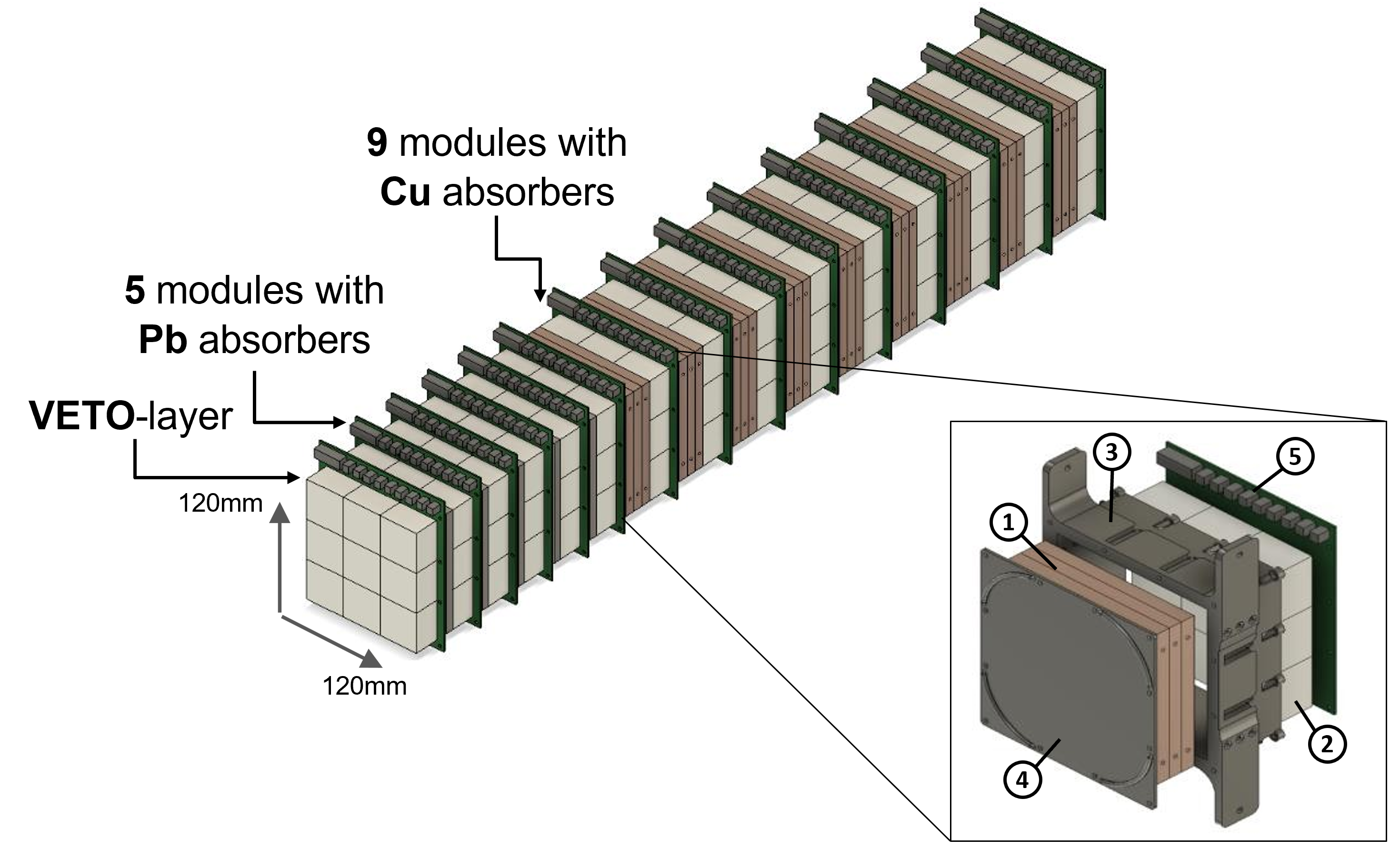}
\caption{Schematic of the HGND prototype. The structure of a single module is shown in the insert. The absorber (1) and nine scintillator cells (2) are placed in light-tight box (3) which is closed from one side with light-tight cover (4). The PCB (5) with nine SiPMs, amplifiers, temperature sensor and connectors is attached directly to the scintillator array.\label{fig:1}}
\end{figure*}

To validate the concept of a full-scale HGND, a small HGND prototype was designed and constructed.
The prototype consists of 15 scintillator layers, sequentially arranged one after another with absorber layers in between~\cite{bib:Zubankov2024}, see Fig.~\ref{fig:1}). 
The first scintillator layer (VETO-layer) is used as a VETO for charged particles.

The first five modules of the HGND prototype placed after the VETO-layer have~8 mm thick lead absorbers. 
This part is used to study in detail the influence of the $\gamma$-background on neutron identification and to develop methods for its rejection. 
The total length of this part is about 7.5~$X_0$ radiation lengths, but only 0.42~$\lambda_{\rm int}$ nuclear interaction lengths. 
The remaining part of the HGND prototype consists of nine modules with 30 mm thick copper absorbers and has a total interaction length of about 2 $\lambda_{\rm int}$.

The transverse size of the detector is $12\times 12$~cm$^{3}$.
The total length of the HGND prototype is 82.5~cm, which corresponds to 2.5~$\lambda_{\rm int}$ nuclear interaction length.
The fully assembled HGND prototype, weighing approximately 100~kg, was mounted on a specially designed and manufactured wheeled frame at a height of 215~cm from the floor level in the plane of the beam axis.
This allows the HGND prototype to be moved to different positions and angles relative to the beam axis.

Each of the scintillator layers consists of a 3 $\times$ 3 matrix of individual scintillators with a transverse area of 4~$\times$~4~cm$^{2}$ and a thickness of 2.5~cm.
Light from each of the nine scintillators is read out by an individual SiPM (Hamamatsu S13360-6050PE)~\cite{bib:Hamamatsu2025mppc} with a sensitive area of 6 $\times$ 6 mm$^{2}$, mounted on a four-layer PCB with readout electronics, necessary connectors and a sensor for temperature correction. These SiPMs were used in the HGND prototype due to the limited availability of EQR15-11-6060D-S units at the time of prototype construction. However, the full-scale HGND will adopt EQR15-11-6060D-S SiPMs due to their lower cost and superior timing resolution, as demonstrated in electron-beam tests~\cite{bib:Guber2024b}.
Each of the nine readout channels consists of a photodetector, an input amplifier, a buffer amplifier with 50~$\Omega$ line capability. Every PCB has voltage regulators for the positive and negative supply for the amplifiers. 
This board is attached directly to the scintillators in the layer.
The absorber, scintillators and the PCB with photodetectors and electronics are packaged in a light-protective module made by 3D printing.

The bias voltage to the modules is supplied by a power supply system consisting of a main power supply with remote digital control and a cascade of boards on a loop cable. 
Each board contains a set of DACs that correct the total bias voltage in the channels and a microcontroller that controls the DACs and implements the temperature correction based on the temperature sensor readings. 
After the amplifiers, the signals are transmitted via 2~meter cables to the TQDC~\cite{bib:AFI2025tqdc}, included in the main data acquisition system.

One of the most important characteristics of a time-of-flight neutron detector is its time resolution. 
The time resolution of scintillation cells of the HGND prototype measured with cosmic muons was 200~$\pm$ 4~ps \cite{bib:Guber2023}.
Taking into account the resolution of the start trigger, the estimated time resolution of the HGND prototype in the experiment is about 270~ps.

\section{Measurements of neutrons from collisions of 3.8\texorpdfstring{$A$}{A}~GeV \texorpdfstring{$^{124}$Xe}{124Xe} with CsI target\label{sec:analysis}}

\begin{figure*}[htb!]
\centering
\includegraphics[width=.605\textwidth]{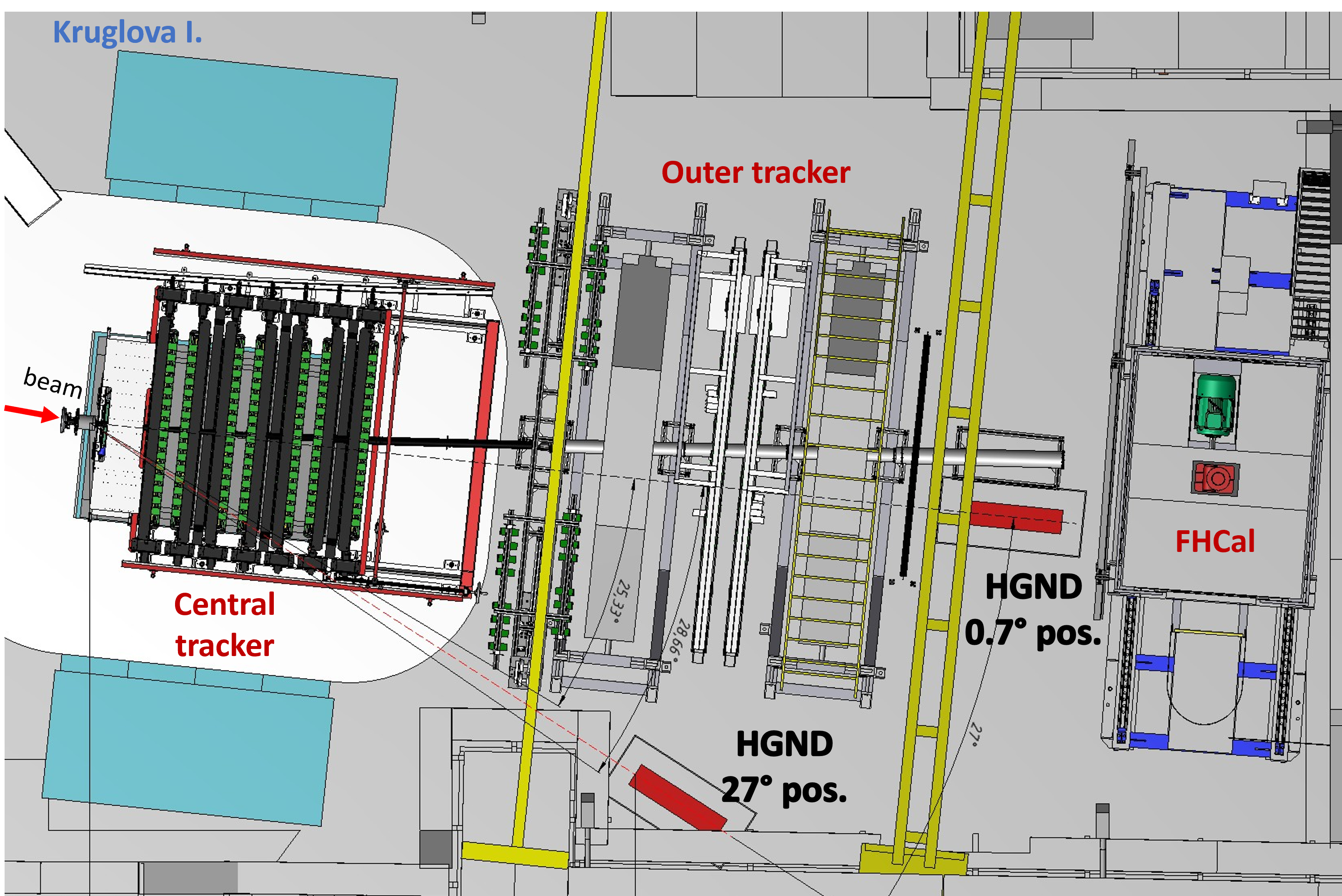}
\quad
\includegraphics[width=.295\textwidth]{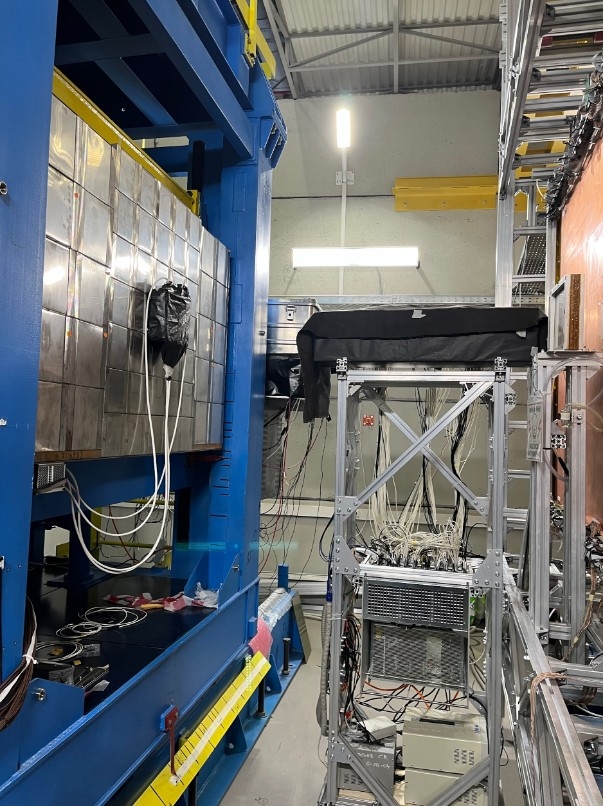}
\caption{Two main positions of the HGND prototype in the BM@N setup (left) and the photo of the HGND prototype at $0.7^{\circ}$ relative to the initial beam axis at distance of 8.35~m from the target (right).\label{fig:2}}
\end{figure*}

The HGND prototype was tested in late 2022 and early 2023 in a physics run of the BM@N experiment to study $^{124}$Xe--CsI collisions at a kinetic energy of the xenon projectile of 3.8$A$~GeV.
The CsI target with a thickness of 1.75 mm, corresponding to a nuclear interaction probability of 2\%, is installed inside the SP-41 analyzing magnet (near its entrance), and the initial ion beam is slightly deflected by the magnetic field (to about $0.7^{\circ}$) relative to its initial direction.
Therefore, the prototype was used in two main positions: at $0.7^{\circ}$ relative to the undeflected primary beam at a distance of 8.35~m from the target to test and calibrate the detector with known spectator neutron energy; and at $27^{\circ}$ relative to the beam at a distance of 5.89~m from the target to measure the neutron spectrum in the mid-rapidity region, see Fig.~\ref{fig:2}.

In this work, the performance of the HGND prototype of detecting forward neutrons from the nuclear fragmentation and EMD of $^{124}$Xe projectiles in interactions with target nuclei was investigated by placing the detector at the $0.7^{\circ}$ position with respect to the initial $^{124}$Xe beam.

\subsection{Selection criteria for hadronic interaction events and EMD leading to neutron emission}
\label{subsec:select}

There are several triggers for hadronic interactions in the BM@N experiment~\cite{bib:Afanasiev2024}.
The Beam Trigger (BT) is generated by the coincidence of the signals of the two Beam Counters (BC1, BC2) and the absence of a signal from the Veto Counter (VC).
The Central Collisions Trigger (CCT) is used to analyze hadronic interactions corresponding to central and semi-central collision events with centrality from 0 to about 60\%.
It consists of the BT trigger, the signal from the Fragment Detector (FD), required to be below the threshold corresponding to non-interacted beam nuclei, and the signal from the Barrel Detector (BD), which is generated when the total multiplicity of hits is greater than four.
The implementation of trigger logic is listed in Table~\ref{tab:1}.
In our analysis of hadronic interactions, a CCT trigger is used with a condition of a single initial $^{124}$Xe nucleus registered by the BC1 beam counter.

\begin{table}[!htb]
\caption{Trigger logic implementation.}
\label{tab:1}
\begin{tabular*}{6cm} {@{\extracolsep{\fill} } lr}
\toprule
\addlinespace[2pt]
BT & $\rm{BC1} \times \rm{BC2} \times \overline{\rm VC}$ \\
\midrule
\addlinespace[2pt]
CCT & $ \rm{BT} \times \overline{\rm FD} \times {\rm BD}$$(\geq4)$ \\
\bottomrule
\end{tabular*}
\end{table}

\begin{figure*}[htb!]
\centering
\includegraphics[width=.9\textwidth]{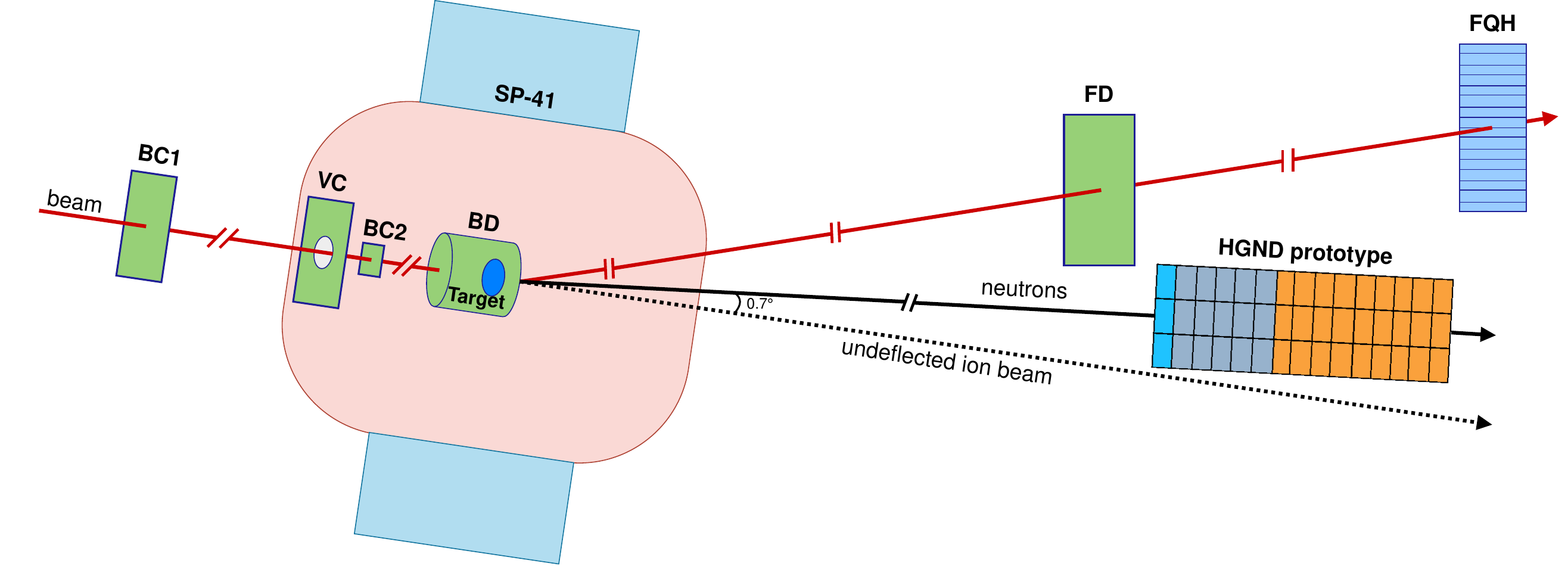}
\caption{Layout of the trigger detectors used for event selection in the BM@N experiment. The Beam Counters (BC1, BC2) and Veto Counter (VC) form a Beam Trigger (BT). Central Collisions Trigger (CCT) is formed by BT and the selection in the Barrel Detector (BD) and the Fragment Detector (FD). The SP-41 analyzing magnet deflects the incoming ion beam, which after the magnet passes through the FD and Forward Quartz Hodoscope (FQH). Since the ion beam is deflected by the SP-41, spectator neutrons from hadronic interactions of beam nuclei and also from their EMD are emitted at $\sim 0.7^{\circ}$ relative to the direction of the undeflected ion beam.\label{fig:3}}
\end{figure*}

The layout of the trigger detectors used for event selection in the BM@N experiment is presented in Fig.~\ref{fig:3}.

However, a dedicated trigger for selecting EMD events is not available in the BM@N experiment.
According to the RELDIS model used to simulate ultraperipheral $^{124}$Xe--$^{130}$Xe collisions, in these EMD events one or two neutrons are emitted by a $^{124}$Xe beam nucleus, without a violent  fragmentation of the projectile~\cite{bib:Pshenichnov2024}. 
Since in addition to forward neutron(s) a single residual nucleus (either $^{122}$Xe or $^{123}$Xe) is produced in EMD without any other particles, the BT trigger with a single initial $^{124}$Xe nucleus in the BC1 beam counter was used to count EMD events to ensure that there was a beam ion entering the target.

\begin{figure*}[htb!]
\centering
\includegraphics[width=.45\textwidth]{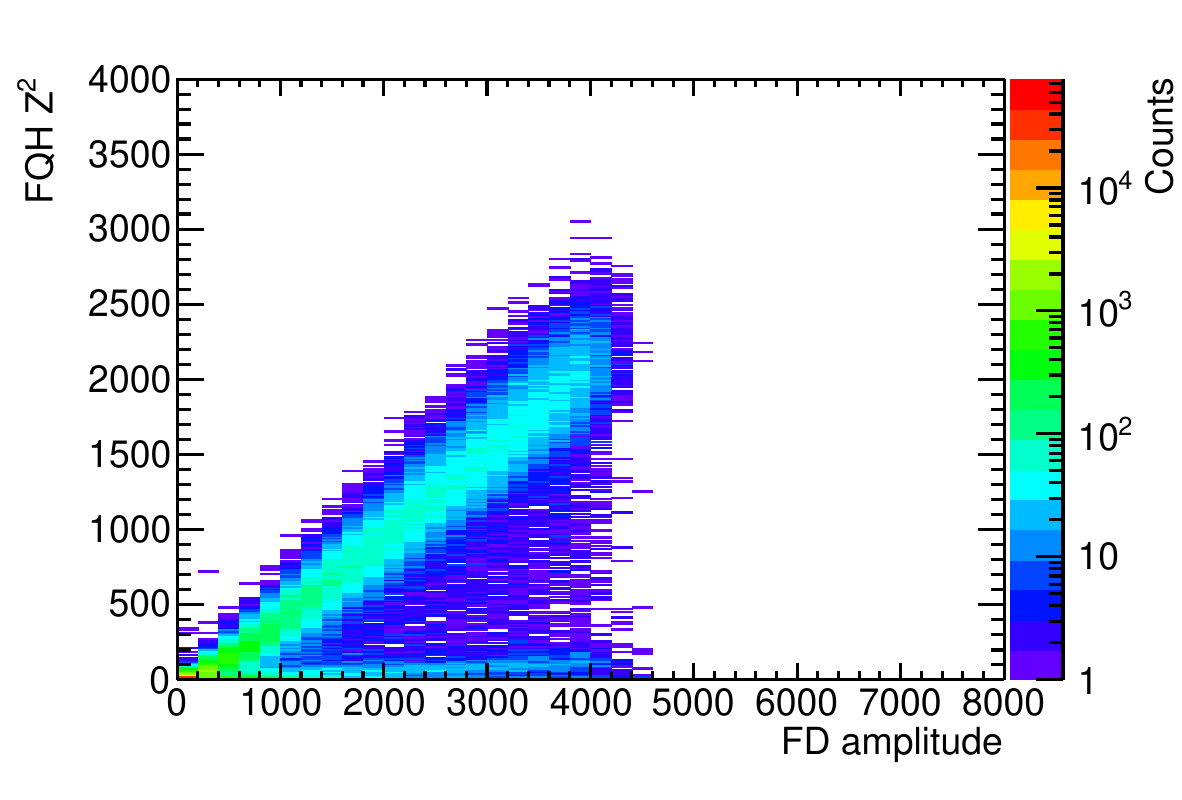}
\quad
\includegraphics[width=.45\textwidth]{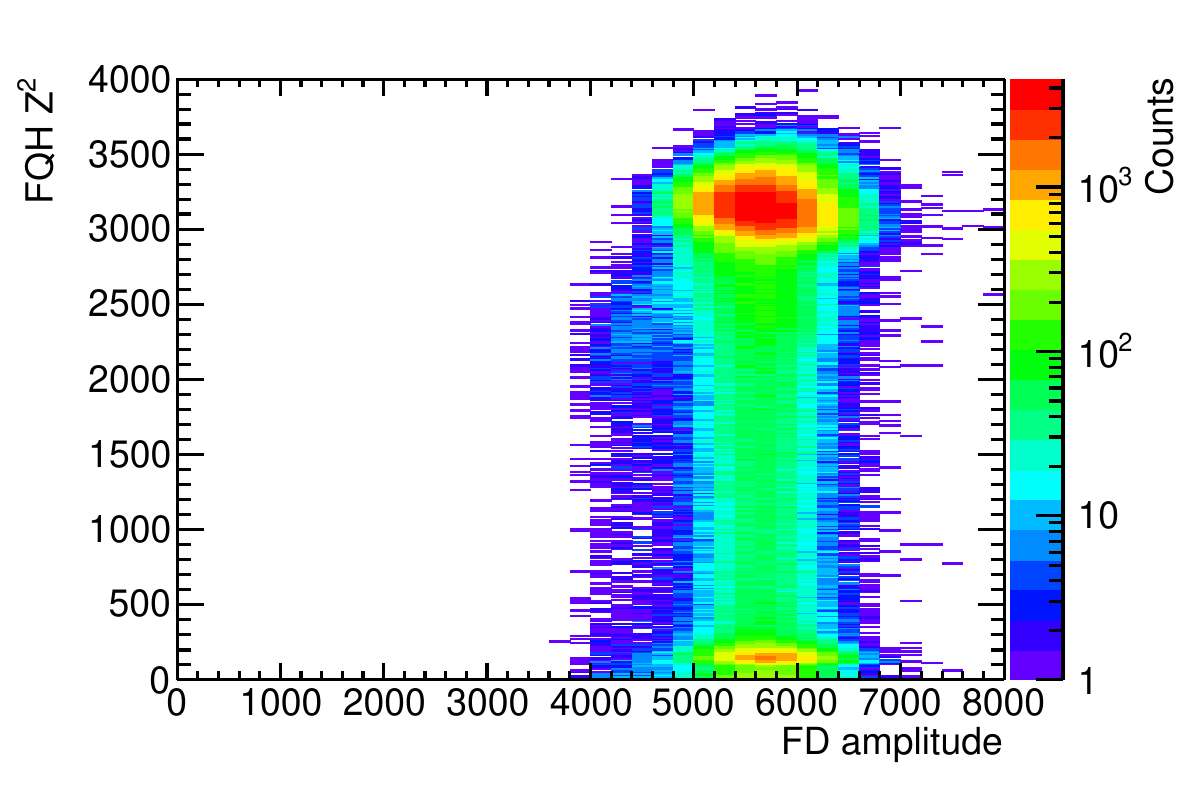}
\caption{Correlation between the $Z^2$ signal in FQH and the FD amplitude for hadronic interactions (left) and EMD events (right).\label{fig:4}}
\end{figure*}

The correlations between the signals from the Forward Quartz Hodoscope (FQH)~\cite{bib:Volkov2023} and the FD after selecting hadronic fragmentation events by the CCT trigger with vertex reconstruction and after selecting EMD events are shown, respectively, in the left and right panels of Fig.~\ref{fig:4}.
The responses of both detectors are defined by the charges of fragments which hit these detectors, and their responses correlate well with each other. As can be also seen, the events with non-interacting $^{124}$Xe nuclei are effectively suppressed, in contrast to EMD events. The domain characterized by high amplitudes in FD and low $Z^2$ in FQH corresponds to events with fragments interacting on their way from FD to FQH or escaping the acceptance of FQH.
A residual nucleus produced in an EMD event is deflected by the magnetic field and hits the FD and the FQH making possible to measure the charge of this nucleus.
A large number of EMD events characterized by the response of the BM@N detectors to residual $^{122,123}$Xe nuclei with $Z^2=54^2=2916$ can be seen in Fig.~\ref{fig:4}, right. 
An additional condition of $Z^2>2500$ in the FQH was imposed to refine the selection of EMD events.

The analysis of the experimental data collected from the HGND prototype starts from the event selection procedure, which includes charged particles rejection, photon rejection, selecting events with two or more triggered cells, amplitude and time-of-flight cuts.
The VETO-layer selection is necessary to distinguish between charged and neutral particles.
If the deposited energy in any cell of the VETO-layer is larger than half of the deposited energy produced by minimum ionizing particles (MIPs), such an event will be discarded.

The photon rejection was used to separate neutrons from photons. 
This procedure rejects an event if any cell in the layer adjacent to the VETO layer is triggered by a signal greater than 0.5 MIP.
Our estimations show that more than 79\% photons and only 10\% neutrons at the entrance to the HGND prototype are rejected with this cut, because of significant difference between the radiation length and the nuclear interaction length in the first lead plate after the VETO detector.

The amplitude and time-of-flight selection, combined with the condition of more than one triggered cell in an event, were used to distinguish  spectator neutrons from a low-energy background.
The selection of successive cells with their amplitudes larger than 0.5 MIP amplitude was used together with the time selection to eliminate background contribution.
The time-of-flight cut was set individually for each active layer. It corresponded to the minimal neutron kinetic energy of about 1.5~GeV.

The ability of the HGND prototype to detect high-multiplicity neutron events is quite limited due to the small transverse dimensions ($12\times 12$~cm$^2$) of this detector. 
This limitation will be removed in the design of the full-scale HGND, with its much larger dimensions, by considering the spatial positions and time stamps of signals obtained from a larger number of individual cells of the HGND.

Given the aforementioned inability of the HGND prototype to detect multiple neutrons from a nucleus-nucleus collision event, our analysis was limited to the detecting only one neutron per event.
The kinetic energy of the neutron was measured using the time-of-flight method by selecting the cell triggered by the particle with the highest velocity in the event.
The measured spectra for both kinds of interactions were divided by the number of incident ions obtained with the BT trigger together with the detection of a single initial $^{124}$Xe nucleus in the BC1 beam counter.

\subsection{Reconstruction of neutron kinetic energy}
\label{subsec:yexp}

\begin{figure*}[htb!]
\centering
\includegraphics[width=.9\textwidth]{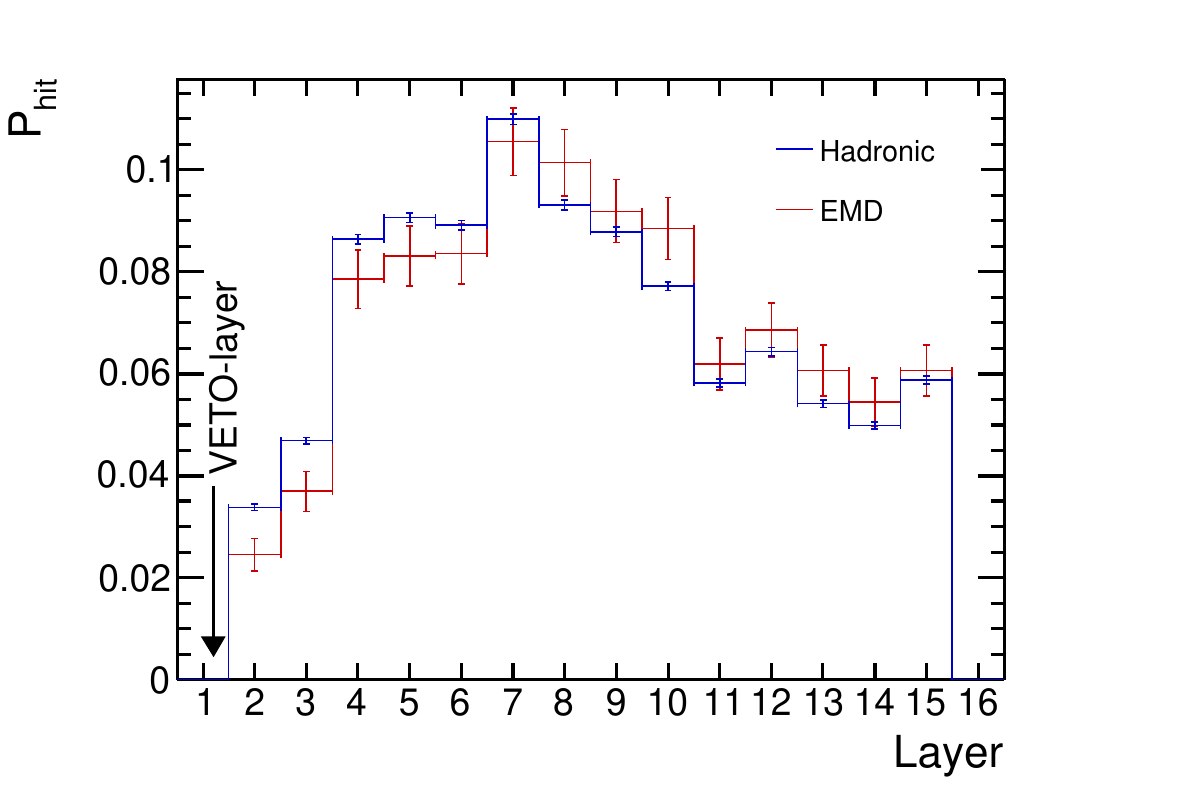}
\caption{Probability distribution to trigger a given scintillator layer of the HGND prototype by particles with the highest velocity in a hadronic or EMD event. This distribution was obtained as a function of the layer number after rejecting photon hits.\label{fig:5}}
\end{figure*}

The measured probability distribution to trigger a given scintillator layer by particles with the highest velocity in the event is presented in Fig.~\ref{fig:5} as a function of the layer number.
As can be seen, neutrons emitted both in hadronic collisions and EMD interact mostly after the 7th layer. 
Photons which hit the HGND prototype were rejected by imposing the condition that if the first layer after the VETO layer is triggered, the event is discarded as associated with a photon.
Due to such a selection, about 20\% of the events of hadronic interactions as well as of EMD were rejected.
The uncertainty of the photon rejection procedure is expected to be one of the main sources of systematic uncertainties in measuring neutron yields.
It was estimated from the change in neutron yield obtained without photon rejection with respect to the standard reconstruction.
Other contributions to the systematic uncertainty of the measurements of neutron yield and neutron flow in the BM@N experiment were not detailed in this work.
This will be a subject of future studies.

\begin{figure*}[htb!]
\centering
\includegraphics[width=.45\textwidth]{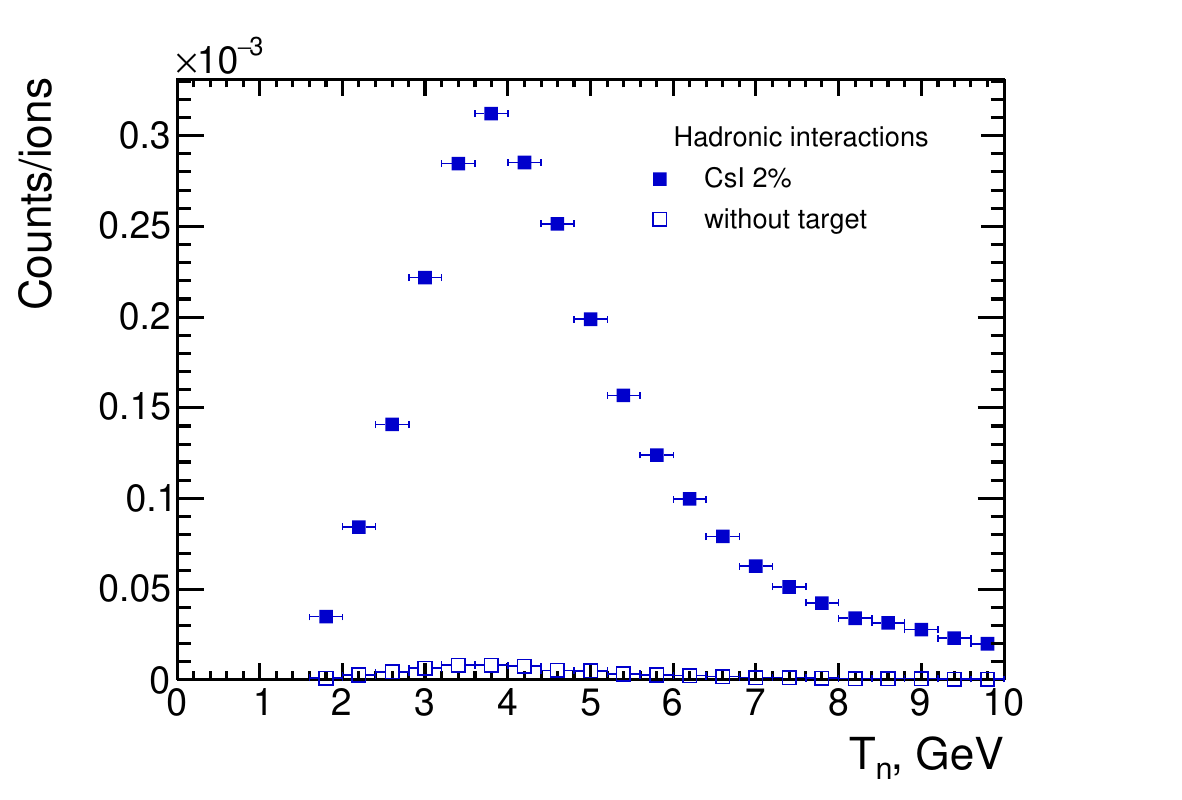}
\quad
\includegraphics[width=.45\textwidth]{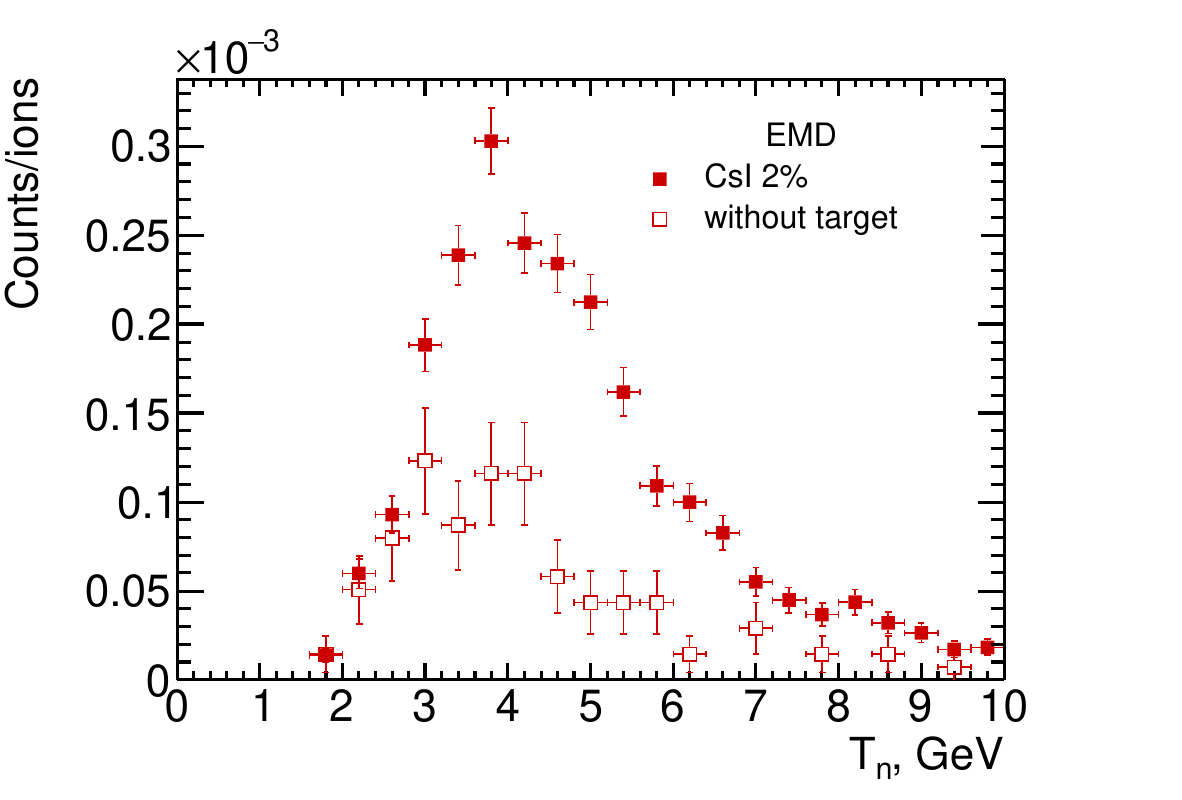}
\caption{Reconstructed neutron spectra with the CsI 2\% target and without it for hadronic interactions (left) and EMD (right).\label{fig:6}}
\end{figure*}

In order to estimate the background caused by the interaction of the beam particles with structural elements of the target station, the number of neutrons produced in a run with the CsI 2\% target was compared with the number of neutrons measured in an empty target run, when the target was removed and the $^{124}$Xe beam interacted only with air or materials supporting the target.
Reconstructed neutron spectra measured with the CsI 2\% target and without it are shown in Fig.~\ref{fig:6} for hadronic and EMD events.
Here, neutron spectra for both hadronic interactions (left) and EMD (right) are normalized to the number of incident ions measured by BT.
The trigger efficiency is close to 100\% for the most central events and slightly decreases for semi-central events in the centrality class 0--60\% selected by the CCT trigger.
Therefore, the effect of trigger efficiency is not considered in this analysis.

The number of neutrons detected in the empty target run is about 3\% for hadronic interactions and 38\% for EMD with respect to neutrons detected in the CsI target run.
Since the procedure of selecting hadronic interactions of beam nuclei with the CsI target in the BM@N experiment is very efficient, the background contribution from the empty target is relatively small for hadronic interactions.
In contrast, because a specific trigger for EMD events is not available, the contribution of background events is noticeably larger in EMD. 
The background contributions described above, mostly associated with neutrons produced in the structural elements of the target station, were subtracted from the measured neutron energy spectra for both interaction types.

\section{Comparison with Monte Carlo simulations}
\label{sec:sim}

\begin{figure}[htb!]
\centering
\includegraphics[width=.9\hsize]{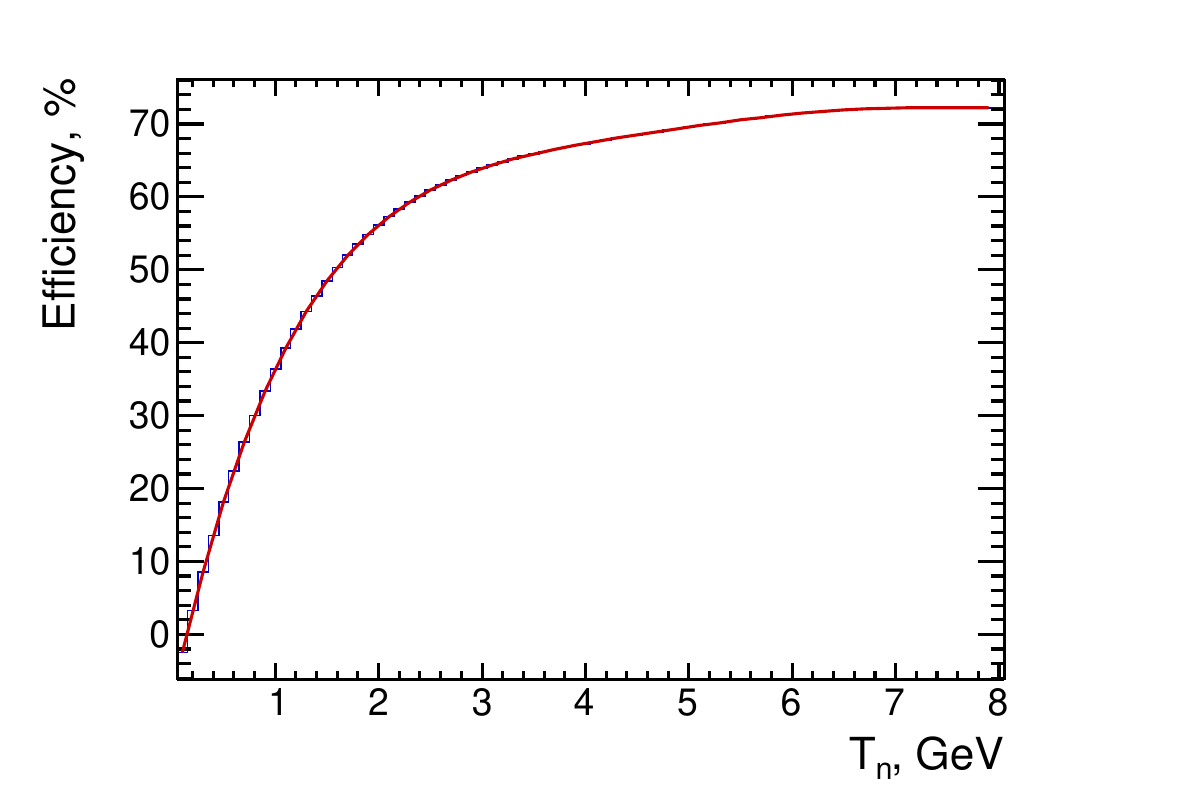}
\caption{Simulated efficiency of the HGND prototype (histogram) to detect a single neutron as a function of neutron kinetic energy obtained with the rejection of charged particles, photons and events with less than two triggered cells, after applying amplitude and time-of-flight cuts. A polynomial fit to the histogram is shown by the red solid line.\label{fig:7}}
\end{figure}

The selection procedure applied to the detected events and described in Sec.~\ref{subsec:select} was used also in modeling.
The efficiency calculated with Geant4~\cite{bib:Geant4} for the detection of a single neutron of a given energy in the HGND prototype is shown in Fig.~\ref{fig:7}.
In this modeling neutrons were transported directly to the front surface of the detector placed in vacuum and were not accompanied by any other particles. 
As can be seen, due to more developed hadronic showers induced by more energetic neutrons, the detection efficiency is higher at higher energies.
A polynomial fit to the modeled efficiency, also shown in Fig.~\ref{fig:7}, was used to correct the number of reconstructed neutrons depending on their energy. 

The hadronic interactions of 3.8A~GeV $^{131}$Xe projectiles with the $^{133}$Cs target as an equivalent of the CsI target were generated with the DCM-QGSM-SMM~\cite{bib:Baznat2019} model.
The total hadronic cross section corresponding to the event centrality of 0-60\% was calculated with this model as 3.165~b.

The RELDIS model~\cite{bib:Pshenichnov2011,bib:Pshenichnov2024} was used to simulate the electromagnetic dissociation of 3.8A~GeV $^{124}$Xe projectiles in UPC with $^{130}$Xe representing the average nucleus in the CsI target.
In this model the Lorentz-contracted Coulomb fields of nuclei in their UPC are represented by equivalent photons, which predominantly excite giant dipole resonances (GDRs) in these nuclei by delivering the excitation energy to  $^{124}$Xe mostly below the proton emission threshold. The total EMD cross section of $^{124}$Xe--$^{130}$Xe collisions was calculated as 1.89~b.

The detection of neutrons by the HGND prototype in its forward ($0.7^{\circ}$) position was modeled with the BmnRoot software taking into account the detailed geometry of the BM@N experiment.
In these simulations, the time resolution of the scintillation cells was set to 270~ps.
We considered hadronic collisions only with the centrality of 0-60\% as selected in the experiment by the CCT trigger, and applied the corresponding centrality selection to simulated events.
In particular, events with an impact parameter of less than 9.54 fm were selected from event files generated with the DCM-QGSM-SMM model to represent 0--60\% centrality.
As demonstrated in Ref.~\cite{bib:Segal2023}, the results of centrality estimators based on the multiplicity of charged particles and on signals from forward calorimeters (which are impacted by forward neutrons, among other particles and nuclear fragments) are consistent within their uncertainties.
In particular, all methods associate the interval $[0, b_{max}]$, where $b_{max}$~$\sim$~9$\pm$1~fm, with the centrality range of 0–60\%.
Therefore, the result of Ref.~\cite{bib:Segal2023} obtained for 4A~GeV Xe+CsI collisions justifies the selection of b~$<$~9.54~fm in the DCM-QGSM-SMM model.

\begin{figure*}[htb!]
\centering
\includegraphics[width=.45\textwidth]{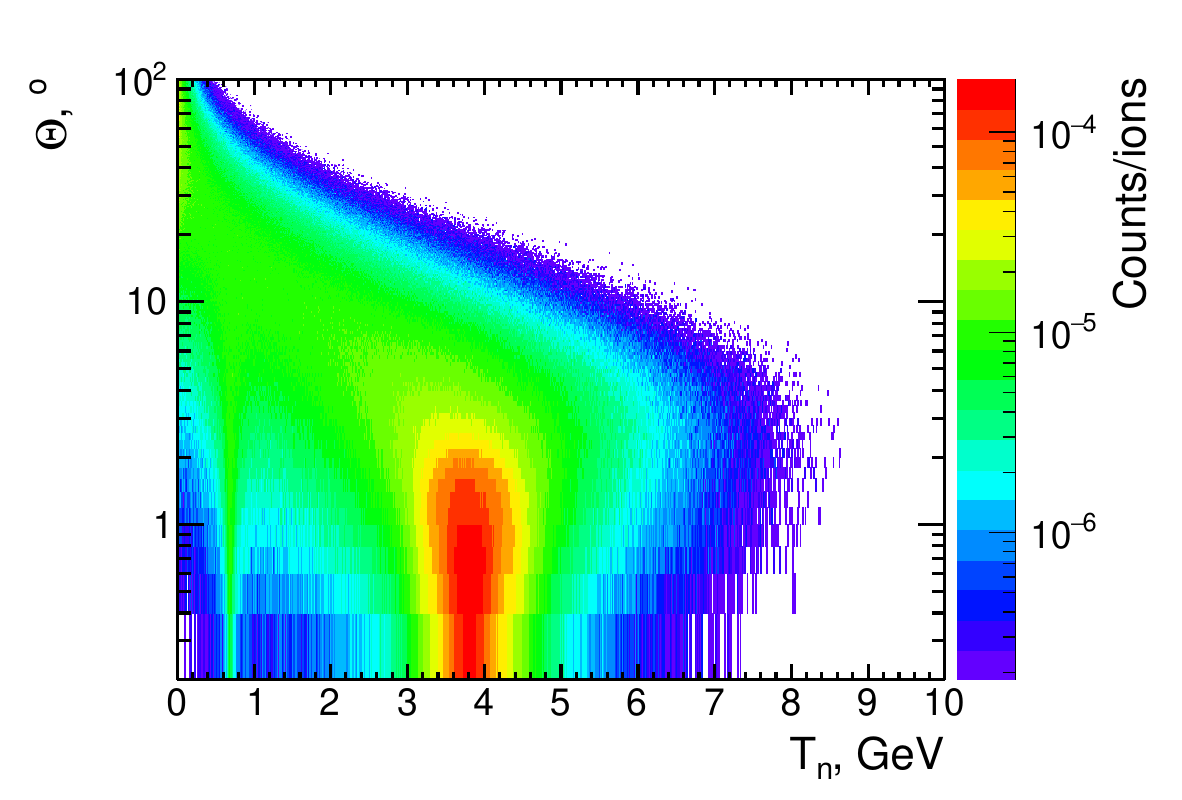}
\quad
\includegraphics[width=.45\textwidth]{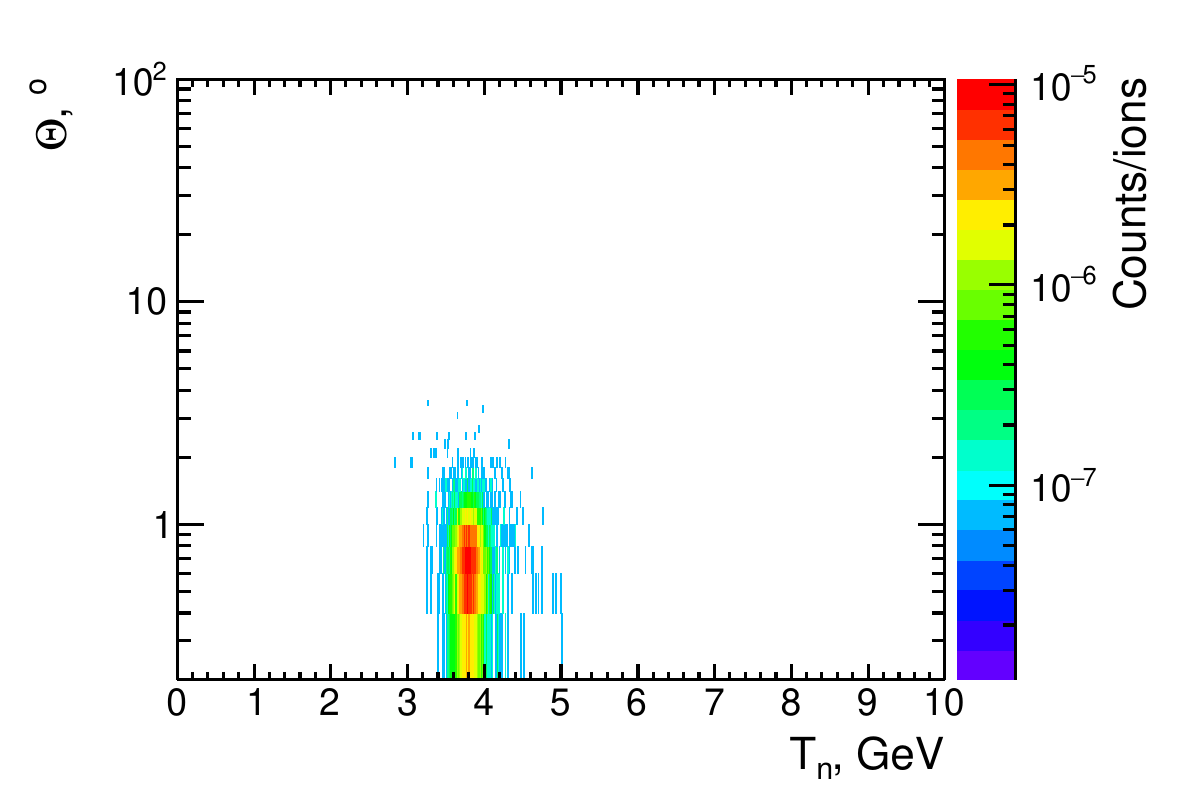}
\caption{Correlations between the emission angle $\Theta$ and kinetic energy $T_{\rm n}$ of neutrons generated in  $^{124}$Xe--CsI collisions. Results of the DCM-QGSM-SMM model (0-60\% centrality, left) and RELDIS model (right) are shown.\label{fig:8}}
\end{figure*}

The correlations between the emission angle $\Theta$ and neutron kinetic energy $T_{\rm n}$ in the laboratory system obtained in simulations with the DCM-QGSM-SMM and RELDIS models are presented in Fig.~\ref{fig:8}.
The spread of kinetic energy $T_{\rm n}$ is essential in hadronic fragmentation events.
For example, few percent of forward neutrons from hadronic fragmentation of $^{124}$Xe have kinetic energy up to 5.5~GeV, i.e., much higher than the nominal beam kinetic energy of $T_{\rm lab} = 3.8$~GeV/nucleon.
In a seminal paper on nuclear fragmentation by A.S.~Goldhaber~\cite{bib:GOLDHABER1974306}, the data on fragment momentum distributions resulting from peripheral nucleus-nucleus collisions were parameterized as $\exp(-p^2/2\sigma^2)$ with a characteristic width of the distribution of $\sigma$ = 90--100~MeV/c estimated for fragmentation nucleons.
A theoretical estimation $\sigma^2=\langle p^2 \rangle/3 = p^2_F/5$ has been obtained in Ref.~\cite{bib:GOLDHABER1974306} by taking randomly an individual nucleon released in fragmentation and considering its intranuclear Fermi motion.
With a typical value of the Fermi momentum of $p_F$ = 230~MeV/c the value of $\sigma \sim$ 100~MeV/c is confirmed, see also Ref.~\cite{bib:PhysRevC.65.051602} which reports $\sigma \sim$~112--116~MeV/c.
It can be expected that following "the $3\sigma$ rule" almost all spectator neutrons have $p<300$~MeV/c ($T<46.8$~MeV) in the center of mass system of the initial $^{124}$Xe nucleus.
In the extreme case of a neutron emitted in the beam direction with $p<~300$~MeV/c in the rest frame of a beam nucleus, the boost with a Lorentz factor of $\gamma$ = 5.0469 provides $T_{\rm n} = 5.52$~GeV in the laboratory system, in a good agreement with the DCM-QGSM-SMM modeling.

As seen in Fig.~\ref{fig:8}, all neutrons in simulated EMD events are characterized by a very narrow angular distribution, and the average neutron energy is equal to the beam energy of 3.8~GeV.
Basically, the energy distribution of EMD neutrons is also very narrow: the calculated energies of all EMD neutrons are between 3.4 and 4.2~GeV, within $\pm 10$\% of the average~\cite{bib:Pshenichnov2024}.
Since the neutrons emitted by electromagnetically excited $^{124}$Xe nuclei mostly result from decays of the low-lying Giant Dipole Resonance in $^{124}$Xe, their kinetic energies in the rest frame of a beam nucleus are below few MeV.

\begin{figure}[htb!]
\centering
\includegraphics[width=.9\hsize]{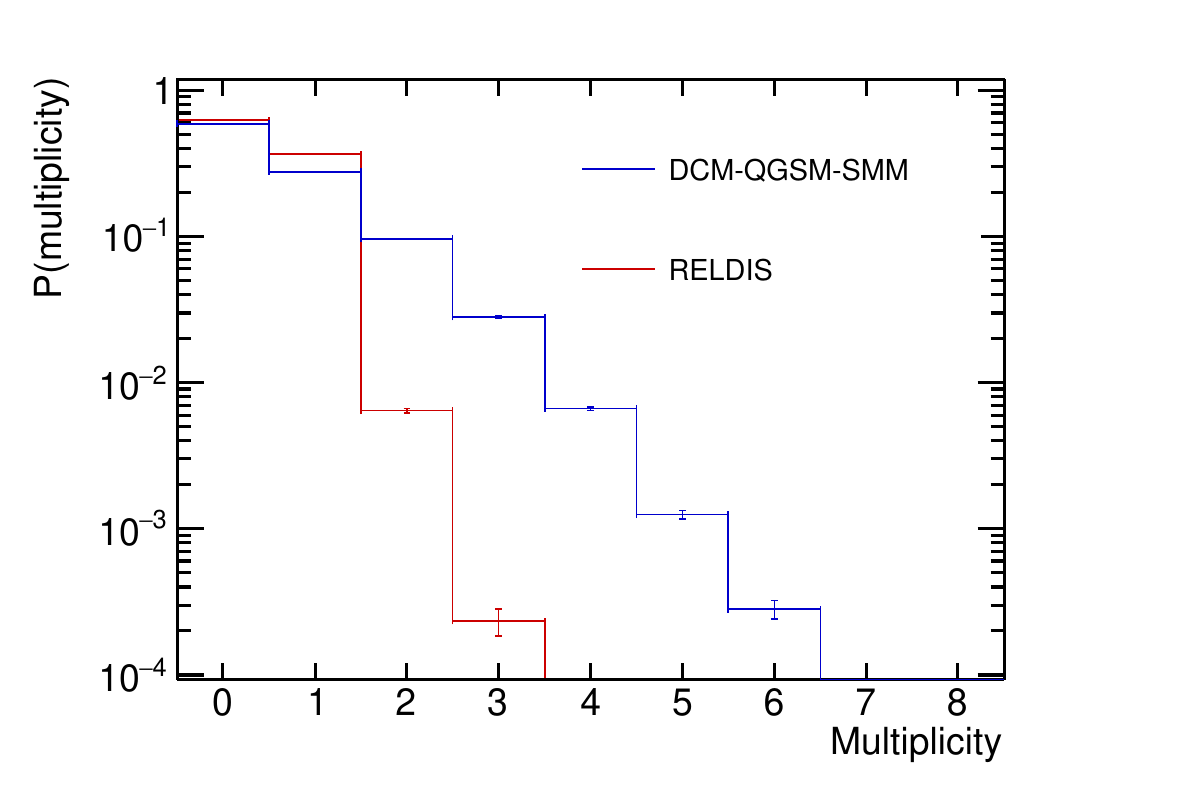}
\caption{Neutron multiplicity distributions on the HGND prototype surface for hadronic interactions and EMD generated with DCM-QGSM-SMM (0-60\% centrality) and RELDIS models, respectively.\label{fig:9}}
\end{figure}

As predicted by RELDIS~\cite{bib:Pshenichnov2024}, the average multiplicity of neutrons produced in the EMD of $^{124}$Xe is 1.05. According to our modeling, because of a limited angular coverage of the HGND prototype, $\sim 66$\% of these neutrons do not hit the detector, see Fig.~\ref{fig:9}. Considering the events only with neutron hits, the average number of neutrons which hit the front surface of the HGND prototype is 1.02. Therefore, the reconstruction of energy of EMD neutrons is free from the above-mentioned complications typical for multineutron events.

In contrast, following the DCM-QGSM-SMM model, the average multitiplicty of spectator neutrons produced in hadronic collisions of the centrality 0-60\% amounts to 16.01 per projectile fragmentation. Due to a much wider angular dispersion of fragmentation neutrons, about 97\% of them are not intercepted by the detector. Again, in those events when neutrons hit the front surface of the detector, their average number is 1.44.

In order to estimate the acceptance and detection efficiency of forward neutrons by the HGND prototype, the same event selection procedures as applied to the experimental data, Sec.~\ref{sec:analysis}, were also used in the modeling.
The acceptance $acc$ and the efficiency $\varepsilon$ of the HGND prototype were calculated on the basis of the above-described modeling using Eqs.~\eqref{eq:acc} and \eqref{eq:eff}, respectively. 
\begin{equation}
\label{eq:acc}
\begin{aligned}
acc &= \frac{N_{\rm{hit}}}{N_{\rm{gen}}} \,,
\end{aligned}
\end{equation}

\begin{equation}
\label{eq:eff}
\begin{aligned}
\varepsilon &= \frac{N_{\rm{rec}}}{N_{\rm{hit}}} \ .
\end{aligned}
\end{equation}
Here $N_{\rm{gen}}$ is the number of neutrons generated in the simulation, $N_{\rm{hit}}$ is the number of neutrons hitting the front surface of the detector, and $N_{\rm{rec}}$ is the number of neutrons reconstructed with the above-described selection criteria.

The calculated values of the acceptance and efficiency of the detection of neutrons from hadronic interaction and EMD are listed in Table~\ref{tab:2}.
\begin{table}[!htb]
\caption{Calculated values of the acceptance and efficiency of the HGND prototype to detect neutrons from hadronic interactions and EMD.}
\label{tab:2}
\begin{tabular*}{8cm} {@{\extracolsep{\fill} } lccc}
\toprule
Model & $acc$, \% & $\varepsilon$, \% & $acc$ $\times$ $\varepsilon$, \%\\
\midrule
DCM-QGSM-SMM & 3.37 $\pm$ 0.02 & 39.69 $\pm$ 0.20 & 1.34 $\pm$ 0.01\\
RELDIS & 36.77 $\pm$ 0.27 & 55.77 $\pm$ 0.41 & 20.51 $\pm$ 0.15\\
\bottomrule
\end{tabular*}
\end{table}

While the HGND prototype is $\sim 50$\% more efficient in detecting more forward-focused EMD neutrons which hit its forward surface with respect to neutrons from hadronic fragmentation, there is a drastic difference in the geometrical acceptance of neutrons from these two processes (3.37\% vs 36.77\%).
Taking into account the resulting $acc\times\varepsilon$, one can conclude that the HGND prototype is much more efficient in detecting neutrons from EMD in comparison to the detection of forward neutrons from hadronic collisions.

It should be noted that the detection efficiency (55.8\%) calculated for EMD neutrons with $\langle T_{\rm n} \rangle =3.8$~GeV using the realistic geometry of the BM@N setup is lower than the efficiency (66\%) calculated for monoenergetic neutrons with $T_{\rm n}=3.8$~GeV  shot directly from a particle source to the front surface of the HGND prototype, see Fig.~\ref{fig:7}. On average, approximately 1.02 neutrons hit the front detector surface per EMD event. Therefore, in some 2\% of EMD events, a pair of neutrons hits the detector surface. Typically, the second neutron is not detected in these events due to the small transverse size of the HGND prototype and the employed reconstruction algorithm. In addition, in the realistic geometry, some neutrons are accompanied by charged particles or photons resulting from the secondary fragmentation of $^{124}$Xe in the structural components of the BM@N setup or in the air. As described in Sec.~\ref{subsec:yexp}, these neutrons are also rejected due to the photon rejection procedure. Taken together, these factors explain the aforementioned difference (55.8\% vs 66\%).

The difference in efficiency values calculated for EMD and hadronic interactions is due to the $\sim 1.5$ times different average multiplicity of neutrons hitting the detector. Since it is impossible to reconstruct more than one neutron in an event with the current detector configuration, the probability of detecting all neutrons in the event is essentially reduced.

The data were normalized to the number of incident ions, for both hadronic interactions and EMD, see Sec.~\ref{sec:analysis}.
In simulations, each generated event corresponds to a single interaction.
For comparison with the measured neutron spectra, the simulated spectra were normalized to the number of incident ions on the target calculated using Eq.~\eqref{eq:i_sim} for both interactions:
\begin{equation}
\label{eq:i_sim}
\begin{aligned}
{N_{ions}} &= \frac{A}{\sigma_{\rm{tot}} \cdot d \cdot N_{\rm{A}} \cdot \rho} \cdot {N_{ev}},
\end{aligned}
\end{equation}
where $A$ is the target molar mass, $\sigma_{\rm{tot}}$ is the total cross section (3.165~b for hadronic interactions from DCM-QGSM-SMM and 1.89~b for EMD from RELDIS), $d$ is the target thickness, $\rho$ is the target density, $N_{\rm{A}}$ is the Avogadro constant, ${N_{ev}}$ is the number of generated events.

\begin{figure*}[htb!]
\centering
\includegraphics[width=.45\textwidth]{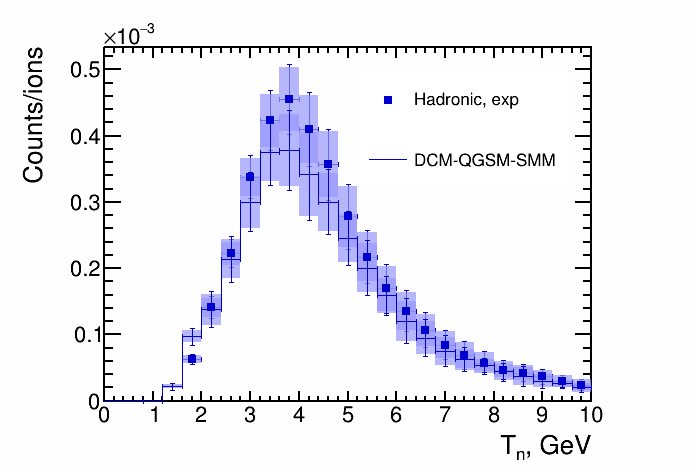}
\quad
\includegraphics[width=.45\textwidth]{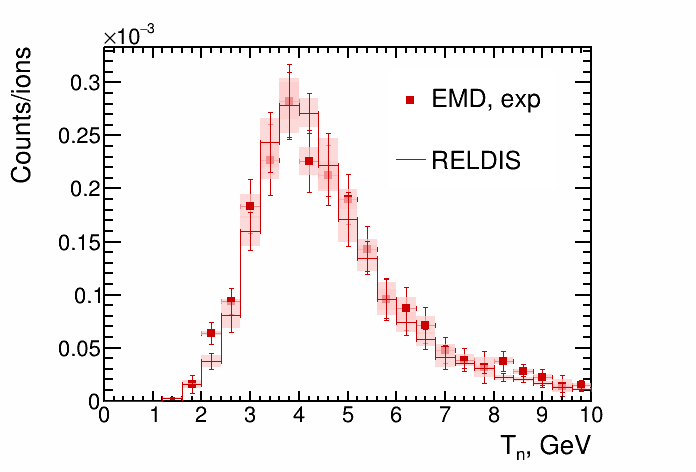}
\caption{Simulated (histograms) and measured (points)  distributions of kinetic energy of neutrons from hadronic fragmentation (left) and EMD (right) of 3.8$A$~GeV $^{124}$Xe on the CsI target. Systematic uncertainties of the measurements are represented by dashed boxes. Combined statistical and systematic uncertainties  are shown by error bars. \label{fig:10}}
\end{figure*}

Finally, the neutron energy distributions corrected for the efficiency of the HGND prototype were obtained from simulations and measurements. 
The spectra are presented in Fig.~\ref{fig:10} separately for neutrons from nuclear fragmentation and EMD of $^{124}$Xe.

\begin{figure*}[htb!]
\centering
\includegraphics[width=.45\textwidth]{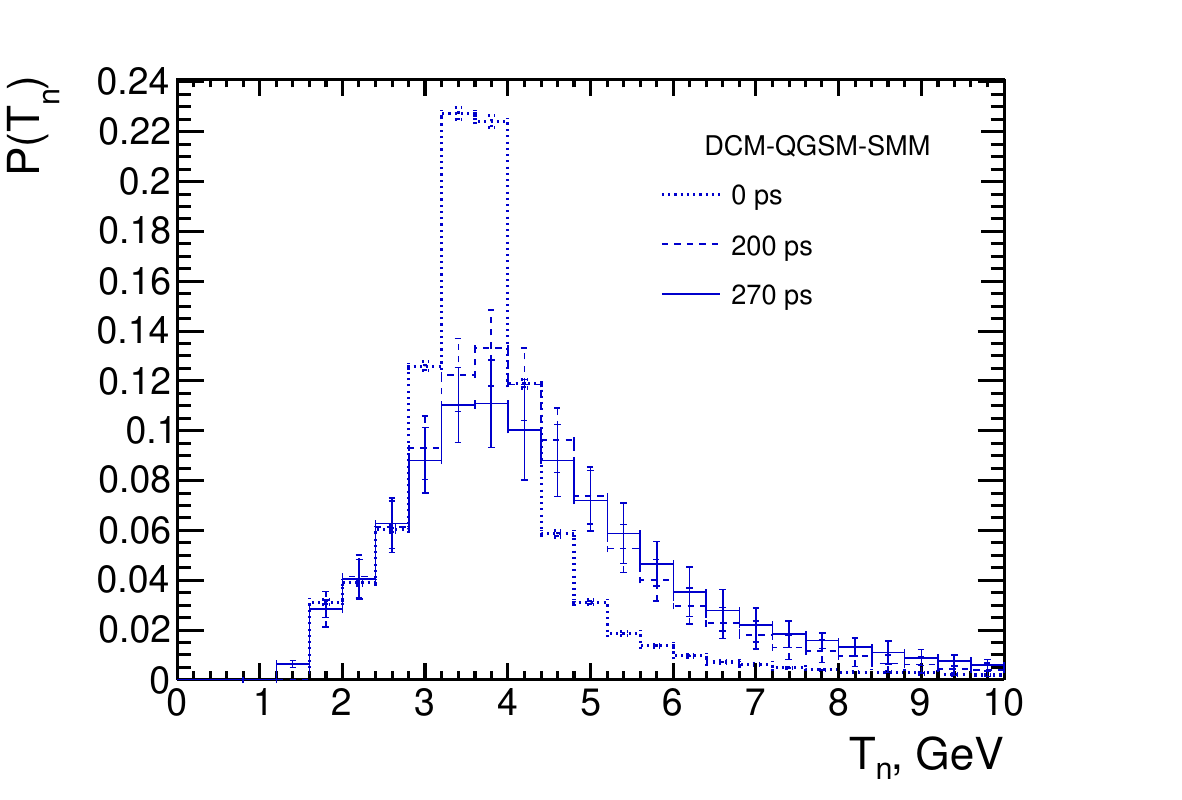}
\quad
\includegraphics[width=.45\textwidth]{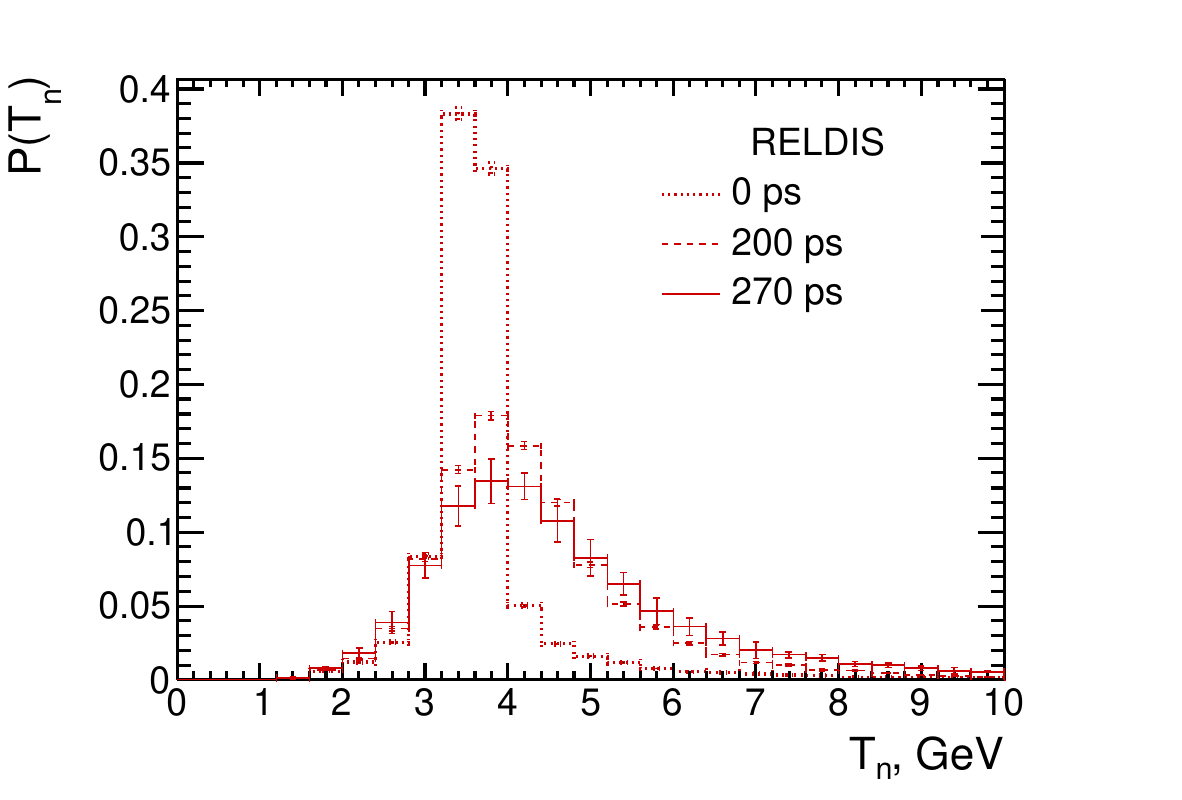}
\caption{Reconstructed kinetic energy spectra of neutrons from hadronic fragmentation (left) and EMD (right) of 3.8A~GeV $^{124}$Xe on CsI target obtained in simulations with three different settings for time resolution: 0, 200 and 270~ps. \label{fig:11}}
\end{figure*}

As can be seen, the measured distributions of the kinetic energy of neutrons agree well with the calculated distributions for both kinds of interactions of $^{124}$Xe.

While the extension of neutron kinetic energy spectra up to 5.5~GeV in hadronic events is the effect of the relativistic kinematics, a further extension of the reconstructed energy spectra up to 10~GeV is caused by a limited resolution of the time-of-flight measurements with the HGND prototype. The neutron kinetic energy spectra reconstructed by the time-of-flight method using the fastest trigger of the scintillation cell are shown in Figs.~\ref{fig:6} and \ref{fig:10}, and they were obtained assuming the time resolution of 270~ps.
In order to demonstrate the sensitivity of reconstruction results to the choice of time resolution, the reconstructed spectra were obtained also with the time resolution parameters of 0 and 200~ps in addition to the nominal 270~ps and they are compared in Fig.~\ref{fig:11}.
In this figure the impact of the time resolution on the shape of the reconstructed spectra is clearly visible.
In the ideal case of 0~ps resolution the initial spectra from event generators were reproduced quite well, but the artifact of the extension of the reconstructed spectra up to 10~GeV is also seen.
One can note that in the present work the extreme case of detecting forward most energetic neutrons has been considered and used for detector calibration.
In future studies of neutron flow in the mid-rapidity region the full-scale HGND will be used to detect neutrons with $T_{\rm n}< 3$~GeV.
As seen in Fig.~\ref{fig:11}, the impact of the time resolution on the reconstructed spectra is essentially alleviated at $T_{\rm n}< 3$~GeV.

\section{Conclusion}
\label{sec:concl}

The design of the HGND prototype, which was built to validate the concept of the full-scale HGND being developed to measure neutron yields in the BM@N experiment at the NICA accelerator complex, was presented. The highly granular multilayer structure of the HGND prototype is represented by interleaved scintillator and absorber layers. The first scintillator layer is used as a VETO to reject charged particles.
As shown by our measurements, forward neutrons emitted in hadronic fragmentation and EMD of 3.8A~GeV $^{124}$Xe nuclei on the CsI target in the BM@N experiment can be identified with the HGND prototype by imposing the photon rejection procedure, proper amplitude and time-of-flight cuts.

The acceptance and detection efficiency of forward neutrons with the HGND prototype were calculated for spectator neutrons from hadronic fragmentation and also for neutrons from EMD of $^{124}$Xe by modeling the transport of neutrons and all other secondary particles in the BM@N setup.
The events of hadronic fragmentation and EMD were generated with the DCM-QGSM-SMM and RELDIS models, respectively.
As found, the simulated kinetic energy distributions of forward neutrons from hadronic fragmentation and EMD of 3.8A~GeV $^{124}$Xe agree well with the measured distributions.

It is shown that the electromagnetic dissociation of relativistic beam nuclei can be considered as a source of well-collimated high-energy neutrons with a multiplicity of about one per EMD event.
It is planned to use such neutrons as well as neutrons from fragmentation of beam nuclei to calibrate the recently developed full-scale HGND~\cite{bib:MOROZOV2025170152}.

\acknowledgments
The authors are grateful to all participants of the BM@N physical run where the data were collected and to the Accelerator Division of the LHEP JINR for providing the xenon ion beam.
Special thanks to Sergey Sedykh (JINR) and Mikhail Kapishin (JINR) for useful discussions and suggestions.

\end{document}